\documentclass{article}
\usepackage{array}
\usepackage{color}
\usepackage{graphicx}
\usepackage{authblk}

\title{Constructing universal Phenomenology for biological cellular systems: An idiosyncratic review on evolutionary dimensional reduction}

\author{Kunihiko Kaneko}
\affil{Niels Bohr Institute, University of Copenhagen, Blegdmsvej 17, 2100 Copenhagen, Denmark, \\
Universal Biology Institute, University of Tokyo, 3-8-1 Komaba, Tokyo 153-8902, Japan
}

\begin{document}
\maketitle
\begin{abstract}
Possibility to establish macroscopic phenomenological theory for biological systems, akin to the well-established framework of thermodynamics, is briefly reviewed. We introduce the concept of an evolutionary fluctuation-response relationship, which highlights a tight correlation between the variance in phenotypic traits caused by genetic mutations and by internal noise. We provide a distribution theory that allows us to derive these relationships, which suggests that the changes in traits resulting from adaptation and evolution are considerably constrained within a lower-dimensional space. We explore the reasons behind this dimensional reduction, focusing on the constraints posed by the requirements for steady growth and robustness achieved through the evolutionary process. We draw support from recent laboratory and numerical experiments to substantiate our claims. Universality of evolutionary dimensional reduction is presented, whereas potential theoretical formulations for it are discussed. We conclude by briefly considering the prospects of establishing a macroscopic framework that characterizes biological robustness and irreversibility in cell differentiation, as well as an ideal cell model. 
\end{abstract}

\section{Possibility for phenomenological theory for biological systems: Can we learn from the success of thermodynamics?}

Biological systems invariably comprise a multitude of components.  Even the simplest cells include thousands of chemical species. The abundances of each component can change over time, and when they increase, cells divide and reproduce themselves.  This prompts a fundamental question: Can we {\sl understand} such dynamical systems of a huge number of chemical species, without going into a detailed description of each molecular process? Note that the hierarchy with diversity in micro-level constituents is not restricted between molecules and cells, but recurs at every scale in biological systems:  In multicellular organisms, developmental process leads to cells of diverse types that coexist and maintain themselves.  In ecosystems, diverse species coexist. In general, it looks hopeless to understand such complex system with diversity, unless some kind of dimensional reduction is available.

In the history of science, however, we find a shining exemplar of success: thermodynamics. Even though the system in concern consists of a huge number of molecules (as we now know), a macroscopic phenomenological theory was formulated by using only a few degrees of freedom, by restricting our concern to stable equilibrium states.  This historical triumph may raise a hope in constructing a phenomenological theory for complex biological systems like cells, consisting of a huge number of chemical species.  

Of course, one could reject such optimism: Unlike typical thermodynamic systems, a biological cell is not in a static equilibrium. Its state changes in time, often with increases in the abundance of chemical components leading to cell growth and division. Moreover, instead of a huge {\sl number} of molecules in thermodynamics, the number of molecule {\sl species} is huge in a cell.  Nonetheless, parallels do exist: biological systems are not at equilibrium, but instead, they frequently attain stationary or steady-growth states. Both thermal equilibrium states and most biological states exhibit stability. In most biological systems, their internal states have tendency to come back to the original after perturbations, as known as adaptation. (Here we are not intending to apply the thermodynamics or statistical physics to biology, but borrow the spirit of it to formulate universal phenomenology for biology). 

Over half a century ago, some biologists began the quest for universal laws and macroscopic descriptions of cellular systems. Monod \cite{Monod} formulated the law of cellular growth, Pirt \cite{Pirt} uncovered a relation between cellular growth rate, cellular yield, and nutrient uptake. Schaechter {\sl et al} uncovered a linear relationship between cellular growth rate and the abundances of ribonucleic acid\cite{Scha}, whereas earlier theoretical attempts for it have been seen \cite{Bremer}.  Waddington \cite{Waddington} introduced his influential epigenetic landscape as a macroscopic potential characterizing the robustness and differentiation of cells throughout development and evolution. At that time, however, quantitative data for cellular states were rather limited.  

Within these few decades, quantitative measurements of cellular states have advanced significantly thanks to techniques like fluorescent imaging, transcriptome analysis, flow cytometry, next-generation sequencing, and so forth. Furthermore, mounting evidence has emerged supporting the characterization of cellular states in terms of a few degrees of freedom. Now, the time is ripe to embark on the quantitative exploration of a general theory bridging the microscopic and macroscopic levels, with statistical physics poised to play a pivotal role. In this context, here, we will survey the possibility of universal theory for cellular states, with recent experimental progresses in mind. 

This proceeding paper gives an idiosyncratic, brief review on the current status of this evolving field termed as {\sl universal biology}. In the subsequent section, we provide a concise overview of the {\sl evolutionary fluctuation response relationship}, which hints at the possibility of representing cells, despite their multitude of components, within a few-dimensional space. We then elucidate how the consistency between the molecular and cellular levels, coupled with the robustness of evolved states, drives a dimensional reduction from a high-dimensional phase space. Furthermore, we show how this evolutionary dimensional reduction imposes constraints and directions on phenotype evolution, which is supported by experimental verification. After discussing the universality of evolutionary dimensional reduction and the potential for related theories, we briefly review several other topics in universal biology that span larger spatial and temporal scales.

\section{Background - why can we expect macroscopic universality in biology?: Evolutionary fluctuation response relationship }

The first insight to the possibility in macroscopic phenomenology to biological system is gained through evolutionary fluctuation response relationship: the phenotypic response throughout the evolutionary course is proportional to the phenotypic variance induced by noise \cite{Sato,KK-ESB}. 

First, recall that the phenotypes, say the concentrations of chemicals, vary over {\sl isogenic} cells that share the same genes.  For statistical physicists, this is not so much surprising as abundances of each component in a cell is not huge, and for cells to grow, components need to be amplified with a positive feedback process. Such isogenic variance of each component, however, is not inherited as in the gene. Hence if such variance is relevant to evolution is not self-evident. 

In relationship, evolution experiment by mutation and selection was recently carried out to enhance a certain property. For instance, by inserting genes to synthesize a protein to generate fluorescence, and by mutating the genes, experiments to select those cells with a higher fluorescence were performed\cite{Sato}. In this case, phenotypic evolution speed was measured as the increase of fluorescence level per each generation upon mutation (genetic change). Here, note that for cells with the same genes, the phenotype, (i.e., the level of fluorescence in this case) fluctuates, as mentioned. By comparing the variance across generations in the evolution, it was found that the evolution speed was positively correlated with the variance.  Furthermore, by using a toy cell model with thousands of components\cite{Zipf,SOC-Zipf}, and making numerical evolution to increase the concentration of a certain component, it was found that the evolution speed was proportional to the variance\cite{KK-ESB,CFKK-JTB}.

The result is reminiscent of the fluctuation-response relation in statistical physics, in the following sense. In this relation, the ratio of the response of a state to an external force is proportional to the variance of the state fluctuation in the absence of the force. In the evolution experiment above, the change in phenotype (e.g. fluorescence) due to the mutation-selection process is proportional (correlated) to the variance before the evolution process. The former may correspond to the response to {\sl generalized, evolutionary force} and the latter to the variance without such a force. Of course, the fluctuation-response relationship in statistical physics is valid in (the vicinity of) thermal equilibrium. Therefore, we cannot apply it directly to evolution. Nevertheless, one may consider the possibility of phenomenological formulation of the evolutionary fluctuation relation, inspired by the above-mentioned similarity.

Recalling linear response theory in statistical physics, a possible extension of the fluctuation-response relationship to phenotypic evolution was proposed as follows\cite{Sato}. Consider a system characterized by a gene parameter $a$ and a phenotypic trait $X$. We can then evaluate the change in $X$ against the change in the value of the gene parameter from $a$ to $a+\Delta a$. 
We then assume the existence of the distribution $P(X;a)$, given by a scalar parameter $a$ for genes. We further assume that it is approximately Gaussian and that the effect of a change in $a$ on the distribution is represented by a bi-linear coupling between $X$ and $a$. (In other words, we choose a variable whose distribution is closer to Gaussian, which can be achieved for example by appropriate variable transformation, while the bi-linear form may be applicable by assuming that the change due to evolution is not so large (this issue will be revisited in later sections)).
With this assumption, the distribution is written by

\begin{equation} P(X; a)= N_0exp(-\frac{(X-X_0)^2}{2\alpha}+v(X,a)).  \end{equation}
with $N_0$ a normalization constant so that $\int P(X:a)dX=1$\footnote{For the lowest order of the change, we neglect the dependence of $\alpha$ on $a$.} . 

Here, $X_0$ is the peak value of the variable $X$ at $a=a_0$, which agrees with the average value of $X$, denoted by $\overline{X(a)}=\int XP(X;a)dX$, whereas the term $v(X,a)$ gives a deviation from the distribution at $a=a_0$. This deviation  $v(X,a)$ can be expanded as $v(X,a)=C(a-a_0)(X-X_0)+...$, with $C$ as a constant, where $...$, a higher order term in $(a-a_0)$ and $(X-X_0)$, which will be neglected in the following analysis.

Assuming this distribution form, $\overline{X(a+\Delta a)}$ is computed from 

\begin{math}
\int X exp(-\frac{(X-X_0-C\alpha \Delta a)^2}{2\alpha}) dX
\end{math}. Then, the change in the average value $\overline{X}$ following the change of the parameter from $a_0$ to $a_0+\Delta a$ is represented by

\begin{equation} 
\frac{\overline{X(a_0+\Delta a)}-\overline{X(a_0)}}{\Delta a}=C\alpha \propto \overline{\delta X^2},
\end{equation}
\noindent by recalling that the variance $\overline{\delta X^2}$ agrees with $\alpha$. 

The above explanation, of course, is not derivation. 
We assumed Gaussian-like distributions\footnotemark, and that the bi-linear coupling term between the parameter and variable, 
which brings about a shift of the average of the corresponding variable.
Empirically, however, several model simulations\cite{CFKK-JTB,KK-ESB} and experiments support this type of evolutionary fluctuation-response relationship\cite{Sato,Ichihashi}.  In this sense, the above distribution provides a phenomenological formulation of the empirical observations, by borrowing the concept in statistical physics. One may also expect the existence of potential function $U(X,a)$ such that $P(X,a) \propto exp (-U(X,a))$, to be discussed later.
\footnotetext{Often the distribution is far from Gaussian. If the variance of the distribution in concern is finite, however,
the distribution can be transformed to be nearly Gaussian, by suitable choice of  transformation of the variable.  For example, if a measured variable $\widehat{x}$  (say the concentration of some protein) follows a log-normal distribution as often observed in biological systems\cite{log-normal}, then,  one can just adopt $X=log(\widehat{x})$ as a phenotype variable in concern. In fact, in most cases log(concentration) or log(fluorescence) is adopted as the variable.}

{\bf Mystery}

{\sl This evolutionary fluctuation-response relationship is associated with the phenotypic variance $V_{ip}$ of isogenic individuals, which is caused by noise. In standard population genetics, in contrast, the phenotypic variance due to genetic variation, named as genetic variance $V_g$ is considered. In fact, the so-called fundamental theorem of natural selection proposed by Fisher \cite{Fisher,Falconer} states that the evolutionary rate is proportional to $V_g$. This relationship is obtained straightforwardly, as the changes in genes are inherited to the offspring, and those from the mother that has a higher fitness for survival can also have a higher fitness\footnotemark .
In contrast, in the aforementioned evolutionary fluctuation relationship concerns with the variance by noise over isogenic cells, and those that happen to take a higher fitness value by noise cannot necessarily produce the offspring with such higher fitness. Hence the evolutionary fluctuation-response relationship is distinct from the Fisher's theorem. If both the relationships are valid, however, this implies that $V_{ip}$ and $V_g$ are proportional throughout the evolutionary course. Both are variance of phenotypes, but the origin for the fluctuation is different: the former is due to noise, the latter by genetic changes. Is the proportionality valid? If so, why can it be?}
\footnotetext{Let us consider the fitness $f$ (degree of adaptiveness of the phenotype, such as the growth rate) and let $P_n(f)$ be the distribution of the fitness of different genotypes at generation $n$. Then the average fitness over genotype distribution is given by
\begin{math}
<f_n>=\int fP_{n}(f)df
\end{math}
Where $<...>$ denotes the average over the distribution $P(f)$. Then the fitness distribution of the next generation is given by
\begin{math}
P_{n+1}(f)=fP_n(f)/\int fP_n(f)df=fP_n(f)/<f_n>
\end{math}
Because the offspring number is proportional to the fitness. From this equation
\begin{math}
<f_{n+1}>-<f_n>= \frac{\int f^2P_n(f)df}{<f_n>}- <f_n> =\frac{(\int f^2P_n(f)df-(\int fP_n(f))^2)}{<f_n>}=\frac{<\delta f_n^2>}{<f_n>},
\end{math}
 where $<\delta f_n^2>=<(f_n-<f_n>)^2>$ is the variance of the fitness $f_n$.
Hence the rate of increase in the fitness per generation is proportional to its variance.}

We first examined this proportionality relationship between these two variances by evolutionary simulations of a catalytic reaction network and gene regulation network models \cite{CFKK-JTB,KK-PLoS}, which demonstrate its validity.
 
Then, to formulate possible relationship between the two variances, a phenomenological theory was proposed. For it, we have revised the previous distribution function 
$P(X;a)$ to include the distribution of genes $a$.  Here the phenotype is (stochastically) determined from the gene $a$, but there is a feedback process from phenotype to genotype through the selection process. Hence, we assume a two-variable distribution $P(X,a)$ both for the phenotype $X$ and genotype $a$.

By using this distribution, $V_{ip}$, variance of $X$ of the distribution for given $a$, can be written as $V_{ip}(a)=\int (X-\overline{X(a)})^2 P(X,a)dX$, where $\overline{X(a)}$ is the average phenotype of a clonal population sharing the genotype $a$, namely $\overline{X(a)}=\int P(X,a)X dX$.
$V_g$ is defined as the variance of the average $\overline{X(a)}$, over genetically heterogeneous individuals and is given by $V_{g}=\int (\overline{X(a)}-<\overline{X}>)^2 p(a)da$, where $p(a)$ is the distribution of genotype $a$ and $<\overline{X}>$ as the average of $\overline{X(a)}$ over all genotypes.

Now we make another assumption on robust and gradual evolutionary process, and introduce {\bf evolutionary stability} hypothesis: Through the course of the (gradual) evolution, the distribution keeps a single peak in the $(X,a)$ space, which leads to Hessian condition for the derivative of $P(X,a)$. It can be easily formulated by assuming the Gaussian distribution for $P(X,a)$ written as:

\begin{equation}
P(X,a)= \widehat{N}\exp[ -\frac{(X-X_0)^2}{2\alpha(a)}+C(X-X_0)(a-a_0)-\frac{1}{2\mu}(a-a_0)^2],
\end{equation}

\noindent
with $\widehat{N}$ as a normalization constant.  The Gaussian distribution $\exp(-\frac{1}{2\mu}(a-a_0)^2)$ represents 
the distribution of genotypes around $a=a_0$, whose variance is (in a suitable unit) the mutation rate $\mu$. 
The coupling term $C(X-X_0)(a-a_0)$ represents the change in the phenotype $X$ by the change in the genotype $a$.  Recalling that the above distribution (5) can be rewritten as

\begin{equation}
\frac{P(x,a)}|{\widehat{N}}=\exp[-\frac{(X-X_0-C(a-a_0) \alpha(a) )^2}{ 2\alpha(a)}+(\frac{C^2\alpha(a)}{2}-\frac{1}{2\mu})(a-a_0)^2]
\end{equation}

\noindent
the average of $X(a)$ for given genotype $a$ satisfies
\begin{equation} 
\overline{X(a)} \equiv \int X P(X,a) dX=X_0+C(a-a_0)\alpha(a).
\end{equation}

\noindent
The evolutionary stability condition mentioned above postulates that the multiplication factor in front of $(a-a_0)^2$ be negative. 
This postulate leads to the condition
\begin{math}
\frac{\alpha C^2}{2}-\frac{1}{2\mu} \leq 0, i.e.,
\end{math}
\begin{equation}
\mu \leq \frac{1}{\alpha C^2} \equiv \mu_{max}.
\end{equation}

\noindent
This means that the mutation rate has an upper bound $\mu_{max}$ beyond which the distribution does not keep a peak in the genotype--phenotype space, and the stability is lost, where the gradual evolution process ceases.
In the above formulation, the distribution is symmetric against the peak, as we assumed the Gaussian-type distribution. However, as the distribution is flatter, the deviation from Gaussian distribution will be amplified. In particular, such assumption with symmetry will be lost. In fact, fitted biological state is rarer than non-fitted one. Then, as the distribution is flatter, non-fitted states are accumulated. In fact, according to the evolution simulation of the catalytic reaction network and gene-regulation network mentioned so far, most phenotypes obtained under large mutation rates have lower fitness\cite{CFKK-JTB,KK-PLoS}, and the evolution to increase the fitness cannot progress when the mutation rate is larger than some threshold. This loss of stability of fitted state under larger mutation rate is analogous to the error catastrophe by Eigen\cite{Eigen}.  The catastrophe here, however, is a result of genotype-phenotype mapping in contrast to combinatorial problems in genetic space in the Eigen's sense.

Now recalling $V_g=<(\overline{X(a)} -X_0)^2>$ and eq.(5), $V_g$ is given by $(C\alpha)^2<(\delta a)^2>$. 
Here, $(\delta a)^2$ is computed by the average $P(X,a)$, as $\mu /(1-\mu C^2\alpha)$.  Then, it follows that

\begin{equation} V_g= \frac{\mu (C \alpha)^2}{1-\mu C^2\alpha} = \alpha \frac{\frac{\mu}{\mu_{max}}}{1-\frac{\mu}{\mu_{max}}}. \end{equation}

\noindent
If the mutation rate $\mu$ is small enough to satisfy  $\mu \ll \mu_{max}$, 
\begin{equation}
V_{g} \sim  \frac{\mu}{\mu _{max}}V_{ip},
\end{equation}
\noindent
is obtained by recalling that $V_{ip}=\alpha$.  Thus, the proportionality between $V_{ip}$ and $V_g$ is obtained.

Intuitive interpretation for this relationship can be speculated as follows \cite{KK-ESB,KK-PLoS}:
In general, the dynamics governing the developmental processes that shape phenotypes are inherently complex, and the final state can be swayed by perturbations both in the initial conditions and those arising during the course of these dynamics. Even if the optimal phenotype is achieved through these dynamic processes, the perturbations introduced by noise during the dynamics can lead to diverse non-optimal states, resulting in the sensitivity of the phenotype to noise.
However, as evolution proceeds, the system gains robustness of the phenotype against the disruptive effects by noise. To achieve robustness to noise, a global attraction towards the desired target phenotype needs to be shaped through the evolution. On the other hand, genetic changes can also introduce perturbations into these dynamic processes. 
If the global attraction to the target state is shaped against the imposed noise, the attraction will not be much damaged, even if the genetic change alters the parameters or equations in the dynamical systems, up to a certain degree. Thus, the increase in robustness to noise is expected to be accompanied by an increase in robustness to genetic changes. 
Indeed, it is observed from numerical simulations that the fitted states become increasingly robust to noise as they evolve, and it is expected that they will also become more robust to genetic changes, leading to a correlation between the two types of robustness.

Now, as the robustness to noise is increased, the phenotypic variance $V_{ip}$ will decrease; similarly, an increase in the robustness to genetic mutation leads to a decrease in $V_g$. Hence, throughout evolution, both $V_{ip}$ and $V_g$ decrease in correlation. This cannot prove the proportionality between the two, but strong correlation between $V_{ip}$ and $V_g$ is expected, as is consistent with the evolution simulations \cite{CFKK-JTB,KK-PLoS} (see also \cite{Ancel-Fontana} for related, pioneering study on RNA, and \cite{Kikuchi,Inoue} for related study on gene-regulation networks).  
For instance, the expression level or concentration of a component may be harder/easier to change by noise if the regulation to maintain the concentration is stronger/weaker. In this case, if a parameter in the system is changed by mutation (e.g. by the change in the enzyme activity for the reaction to the corresponding component), the concentration will also be harder/easier to change. This may provide a possible molecular mechanism for the correlation between phenotypic changes due to noise and mutation.

\section{Restriction to steady growth instead of equilibrium}
{\bf Next mystery} 
{\sl Coming back to the distribution theory of $P(x,a)$ there remains another mystery: Why can we consider just the distribution of few variables? Recall that biological systems involve a huge number of degrees of freedom for the phenotype:  In fact, a bacterial cell contains thousands of components, whereas the models we adopted above involve a thousand of components.  Next, how are genetic changes represented by just one parameter $a$? Generally, genes are represented by a base sequence of nucleic acids in DNA? How can it be represented by a scalar variable\footnote{One possible  candidate for such scalar parameter is Hamming distance from the fittest sequence, or a projection according to the corresponding phenotype \cite{SatoKK2007}.}? Furthermore, how phenotypic and genetic changes which are in a quite different time scale can be treated in the same dimension? In short, how is the description by few degrees of freedoms possible?}

\begin{figure}[ht]
\begin{center}
\includegraphics[width=10cm]{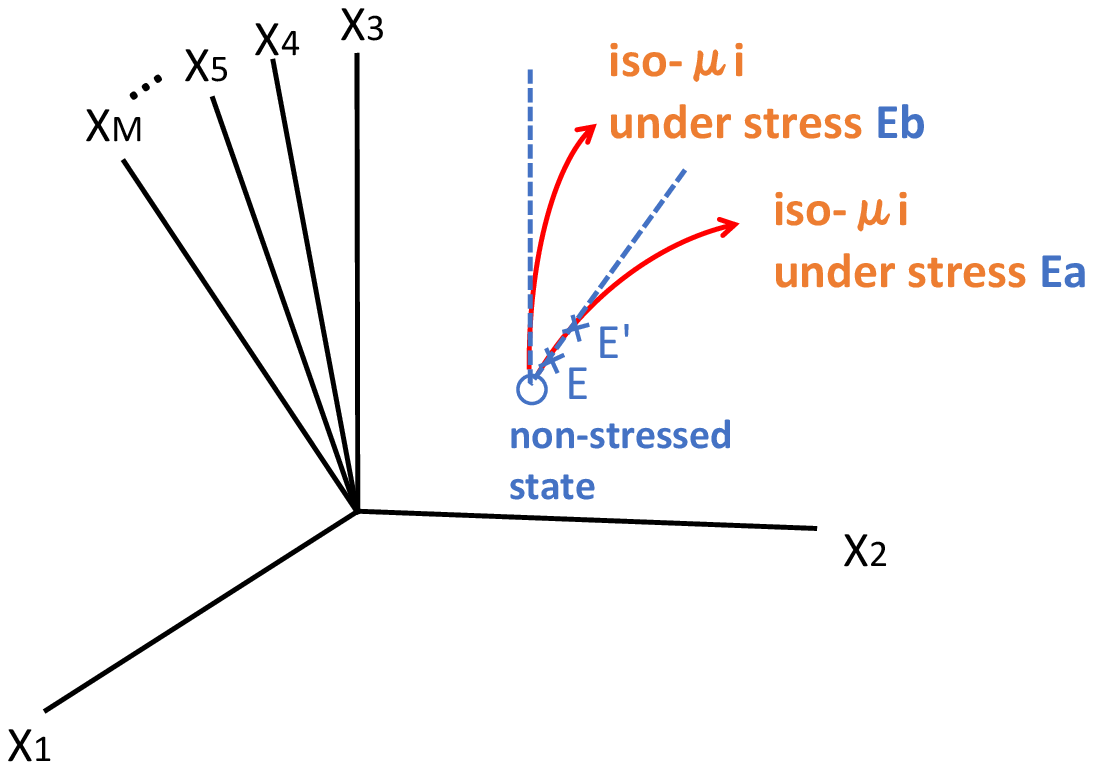}
\end{center}
\caption{Schematic representation in the phenotypic change in a high-dimensional state space. Steady-growth condition leads to the curve where the growth rates of each component are identical.  Changes in gene expression in a high-dimensional state space upon certain perturbations $\delta E$, with the increase in their strength are schematically presented. By increasing the strength of a given environmental stress, the phenotypic changes in component concentrations follow a curve satisfying the constraint that the growth rates of all components are identical, i.e., an iso-$\mu$ line $F_1=F_2=\cdots =F_M$, in an $M$-dimensional state space. 
For a different type of environmental stress, the locus in the state space follows a different iso-$\mu$ line.
}
\end{figure}

One straightforward origin of such dimensional reduction could be due to the condition for the steady state.
By restricting our concern to cells with the steady growth, some phenomenological laws have been uncovered. They cover the classic, empirical relations by Monod\cite{Monod} and Pirt\cite{Pirt} as mentioned above, as well as a relation on the change in protein abundances formulated recently\cite{Scha,Scott}. 

Restriction by the steady growth has recently been formulated generally\cite{Mu}. Consider a cell system with $M$ components, and let $\mu_i$ be the growth rate of component $i=1,2,\cdots,M$. i.e., the abundances $N_i$ of each component increase with $\exp(\mu_i t)$ over a cell division cycle.  Then, the steady-growth implies that abundances of every component have to be approximately doubled for cell division, which implies that $\mu_i (\{N_j\})$ is equal for all components $i$, which coincides with the growth rate rate of cell volume (the whole cell contents) denoted by $\mu_g$.  This leads to $M-1$ constraints over all the concentrations. Then, consider the application of a given environmental stress. After adaptation to the novel environmental condition, the above $M-1$ constraints have to be satisfied, so that the change in cell's state would be restricted along a one-dimensional manifold in the $M$-dimensional compositional space, against the increase in environmental stress (see Fig.1).  

To formulate this constraint for steady-growth, let us denote the concentration of each component $i(=1,\cdots,M)$ by $x_i(>0)$. The cellular state is represented as a point in an $M$-dimensional state space.
Here, each component is synthesized or decomposed relative to other components at a rate $f_i(\{ x_j \})$, for instance, by the rate-equation in chemical kinetics. 
Additionally, all concentrations are diluted by the rate $(1/V)(dV/dt)=\mu_g$, so that the time-change of a concentration is given by
 
\begin{equation}
dx_i/dt=f_i(\{x_j\}) -\mu_g x_i.
\end{equation}
 
For convenience, let us denote $X_i=\log x_i$, and $f_i=x_i F_i$. Then, Eq. (9) can be written as
\begin{equation}
dX_i/dt=F_i(\{X_j\}) -\mu_g,
\end{equation} 
In which it is assumed that $x_i\neq 0$, i.e., all components exist. Based on this, the stationary state is given by the fixed-point solution
\begin{equation}
F_i(\{X^*_j \}) =\mu_g 
\end{equation}
for all $i$.
 
Now, consider intracellular changes in response to environmental changes as represented by a set of continuous parameters $E^a$, the strength of environmental stress of the type $a$. In other words, each environmental change of the stress type $a$ is parameterized by a continuous parameter $E^a$ (such as the temperature, nutrient limitation, etc.).
Using this parameterization, the steady-growth condition leads to
\begin{math}
F_i(\{ X^*_j(E^a) \},E^a)=\mu_g(E^a).
\end{math}

In response to environmental changes, the growth rate $\mu_g$ changes but the above equations have to be satisfied, so that the $M-1$ conditions $F_1=F_2=\cdots =F_M$ must be satisfied.
Thus, a cell must follow a 1-dimensional curve in the $M$-dimensional space (see Fig. 1) under a given change in the environmental conditions (e.g., stress strength).

We then linearize the change around the original (non-stressed state) $E_0$.
With this each $X_j^*$ changes from $X_j^*$ at $E_0$, to $X_j^*+\delta X_j$, accompanied by the change from $\mu_g$ to $\mu_g+\delta \mu_g$. Assuming a gradual change in the dynamics $x_j$, we introduce a partial derivative of $F_i(\{ X^*_j(E) \})$ by $X_j$ at $E=E_0$, which gives the Jacobi matrix $J_{ij}$. With this linear approximation,
we obtain
\begin{equation}
\sum_j J_{ij} \delta X_j (E) + \gamma_i \delta E =\delta \mu_g (E) =c \delta E
\end{equation}
with
\begin{math}
\gamma_i \equiv \frac{\partial F_i}{\partial E},
\end{math}
\noindent where from linear approximation $\delta \mu_g=c \delta E$ with a constant $c$. Accordingly, we obtain
\begin{equation}
\sum_j J_{ij} \delta X_j (E) =\delta\mu_g (E)(1-\gamma_i/c)
\end{equation}
By putting $E$ or $ E'$ in the above equations, and dividing the equation for $E$ by that for $ E'$, we get

\begin{equation}
\frac{\delta X_j(E)}{\delta X_j (E')} =\frac{\delta \mu_g (E)}{\delta \mu _g(E')}
\end{equation}
\noindent across all $j$, against two stress strengths $E$ and $E'$.

In other words, the change in $\delta X_j$ is proportional across all components $j$ with a common ratio.  Now one can examine this relationship by measuring the changes in all the concentrations of mRNAs or proteins by using the transcriptome or proteome analysis. By changing the strength of stresses from the original state to $E$ and $E'$, and $\delta X_j$ was measured across thousands of mRNA species for {\sl E coli}. Interestingly, the relationship (14) is satisfied across most components, even against a large degree of environmental changes\cite{Mu}(based on the data in \cite{Matsumoto}). Indeed, the linear approximation is valid up to the degree at which the growth rate goes down almost to 20 \% from the original, below which the steady growth no longer holds. In other words, the common linear relationship is valid across all components, as long as the steady growth is satisfied.

Interestingly, the relationship (14) is approximately true for most of a few thousand components (mRNAs or proteins), even across {\bf different types} of environmental stresses (i.e., heat, osmotic pressure, starvation, or other chemical stresses) ${\bf E}$\cite{Mu,CFKK-PRE} (as well as in the protein data based on \cite{Heinemann}). This is not explained in the above steady-growth theory, because the directions of the steady-growth curves would be different by different stress types $a$(given by vectors ${\bf E^a}$, as in Fig.1), or in other words, $\gamma_i$ depends on each stress type $a$. This experimental observation, then, suggests that drastic dimensional reduction of phenotypic responses occurs in the living cells, i.e., even across different environmental perturbations (stresses).

\section{Dimensional reduction as a consequence of robustness and plasticity}

{\bf next mystery: {\sl deep linearity}}: {\sl The above proportionality relationship would not be explained just by the steady-growth condition. Indeed, when we studied a toy cell model consisting of a large number of components with arbitrarily chosen reaction dynamics, first the proportionality in eq.(14) was valid only against a tiny change in the environment. Furthermore, if we put different types of environmental changes, eq.(14) would not be valid either. Then, what is the origin of dimensional reduction in a biological system?}

\begin{figure}[ht]
\begin{center}
\includegraphics[width=10cm]{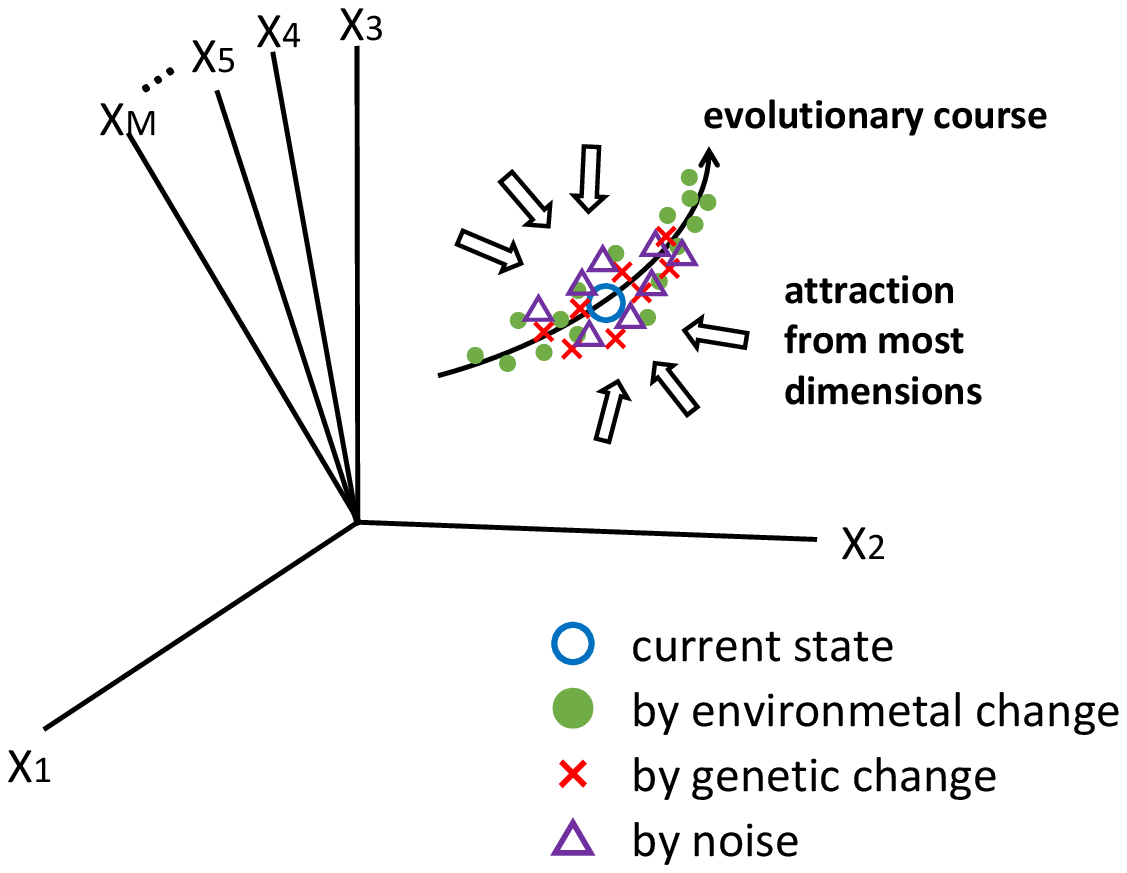}
\end{center}
\caption{
Robustness of phenotype to perturbation and plasticity along the environmental condition leads to dimension reduction against different environmental conditions, genetic (evolutionary) changes and noise.
}
\end{figure}

Here one should recall that a biological system (cell) is shaped as a result of evolution. Such cellular dynamical-system that has a higher fitness (growth-rate) is selected under a given environmental condition. Note such dynamics are under perturbations.  Intracellular reactions are noisy as the number of molecules therein is not so large, and the environmental condition fluctuates. Then the fitted phenotype state should be robust to such perturbations. Recalling that the stability of an equilibrium state allows for thermodynamic description by few macroscopic variables, one may expect that robustness in evolved biological systems similarly allows for reduction to a few macroscopic variables from a high-dimensional phenotypic space.  

In fact, we considered the toy cell model of a huge number of components, in which the abundances of each component change as a result of stochastic intracellular dynamics with catalytic reactions, as well as by taking nutrient chemicals from environment, leading to the cellular growth, as adopted in the study of \S 2\cite{CFKK-JTB}.  The dynamics thus involve a large number of concentration variables, which provide the high-dimensional phenotypic space. The growth rate (fitness) is determined from these phenotypes, and accordingly we evolved the reaction network so that a fitted phenotype (e.g., higher growth) is realized. After the evolution, we found that the relationship (14) is valid across large environmental perturbations and over different types of environmental stresses\cite{CFKK-PRE}.

Now the origin of this {\sl deep linearity} is understood as follows, based on the dynamical-systems analysis.
The shaped fitted state from this evolved network is robust to inevitable perturbations mentioned above\cite{CFKK-PRE}.
Then the fitted state (attractor) has to be attracted from most directions in the high-dimensional state space. In terms of dynamical systems, this means that there is a sufficiently strong flow from most directions to the (fixed-point) attractor corresponding to the fitted state. If this attraction occurred from every direction, however, the evolution to increase the fitness would be difficult, as the state would not be easily changed by slight alteration in dynamical systems (flow in the state space) induced by genetic changes (mutation). Then, along the direction in which evolution has occurred or is ongoing, the phenotypes are feasible to change (see Fig.2 schematically).    All the concentration changes will be constrained along a one-dimensional (or few-dimensional) manifold. If we consider the Jacobi matrix for the relaxation dynamics to the stable state, the magnitude of one (or few) negative eigenvalue(s) is close to zero, whereas those of others are much more negative.

The above picture on evolutionary dimensional reduction has been confirmed numerically by studying the cell model mentioned above \cite{CFKK-PRE,Sato-KK1}. With the numerical evolution, it was found that after evolution, the global proportionality across a variety of environmental conditions is shaped. In this case, changes in most concentration variables induced by a variety of distinct environmental perturbations are found to be highly restricted along a one-dimensional manifold. One of the (negative) eigenvalues in the Jacobi matrix is close to zero, and the principal component matches with its eigenvector.

Now, the response to any environmental perturbations occurs along the above one( or low)-dimensional manifold, and then we do not need to be concerned with all the eigenvalues (and eigenvectors), but only the largest one that dominates the behavior.  Then with straightforward calculation, it is shown that eq.(14) holds across different directions in environmental changes. Also, as the magnitude of the largest eigenvalue is close to zero, the change along the eigenvector is extended, so that the region of the linear manifold is extended.  Thus, the deep linearity is explained by the evolutionary dimensional reduction. 

From the above discussion, it is expected that the change due to genetic variation is also constrained along the same manifold. In fact, the concentration changes $\delta X_i$ by genetic variations, by environmental changes, and by internal noise lie on the same curve, according to the simulation of the above cell models\cite{CFKK-PRE}.  
Accordingly, the responses due to environmental and evolutional (genetic) changes will be correlated, i.e., the relation (14) is satisfied between $\delta X_i$ by the environmental and genetic (evolutional) changes.

Following this argument, consider the change in $\delta X_i$ and $\delta \mu_g$ as a function of environmental change $E$ and genetic change $G$, and then extend the relationship (14) to the case with the genetic evolution after the application of environmental stress. In this case, by representing the changes against $E$ and $G$, we get the relationship
\begin{equation}
\frac{\delta X_i(E,G)}{\delta X_i(E,0)}=\frac{\delta \mu_g(E,G)}{\delta \mu_g (E,0)}.
\end{equation}
Here, by the applied stress, the growth rate is decreased from the original state, so that $\delta \mu_g(E,0)$ is negative, whereas with the evolution, the growth rate is recovered toward the original, so that the change from the original non-stressed state, $|\delta \mu_g(E,G)|$ should be smaller than $|\delta \mu_g(E,0)|$.  This implies that $\delta X_i(E,G)/\delta X_i(E,0) <1$. Hence, as the evolution progresses to recover the growth, the change in each mRNA (or protein) concentration induced by the stress is reduced with the evolution.  Hence the long-term response by evolution occurs to the direction to cancel out the environmentally induced change, as is observed in bacterial evolution experiments\cite{CFKK-Interface,Horinouchi1,Horinouchi2}. This {\sl  homeostasis} by adaptation is akin to Le Chatelier principle in thermodynamics.

To close this section, recall the response and fluctuations are two sides of coin, in statistical physics, whereas that phenotypic fluctuations over isogenic cells ($V_{ip}$) are often large as recent experiments have demonstrated\cite{Elowitz,log-normal}. Now, because the variations $\delta X_i$ by noise and by mutation are also located at the common low-dimensional manifold (Fig.2), it is shown that the variances by internal noise ($V_{ip}(i)$) and by genetic variation ($V_g(i)$) are proportional across components $i$. This $V_{g}(i)-V_{ip}(i)$ law is confirmed in the toy cell models\cite{CFKK-PRE} numerically and is also consistent with some experimental results\cite{Stearns,Landry,Lehner,Uchida}. 

\section{Universality of evolutionary dimensional reduction}

To recapitulate the preceding discussion, we posit the following \textit{evolutionary-dimensional reduction hypothesis}: Over the course of evolution, phenotypic states tend to exhibit robustness, yet they display plasticity in specific dimensions crucial for evolutionary adaptation. Consequently, a limited number of modes, characterized by eigenvalues close to zero (termed slow modes), become isolated from the rest. In its simplest manifestation, this dimensionality reduction reduces to a single dimension. However, in scenarios where multiple survival conditions must be met or when distinct environmental factors exert influence throughout evolution, the dimensionality may encompass several facets.

As most variations arising from noise, environmental alterations, and genetic mutations are constrained primarily along this slow manifold, these changes from the different sources exhibit correlations, as proposed by the evolutionary-fluctuation response relationship. This phenomenon engenders both predictions and constraints in the evolution of phenotypes.

Moreover, the separation of slow modes fosters enhanced evolvability. When all modes share a similar timescale, mutations affecting each mode would be effectively cancelled out and have minimal impact on the phenotypic dynamics due to the law of large numbers. Hence, they yield only slight fitness alterations from random mutations, akin to the proverb, {\sl too many cooks spoil the broth}. Conversely, when few slow modes are separated, most faster modes are controlled by these slow modes (as in the adiabatic elimination of fast variables\cite{Haken,KK-adel}). Consequently, mutational changes induce collective alterations in other variables, thereby amplifying their influence and accelerating the evolution speed.

Given the apparent validity of these premises across a wide array of biological systems, we anticipate that this evolutionary dimensional reduction holds universal significance. In support of this hypothesis, we have conducted evolution simulations under specific fitness conditions and confirmed (a) the evolutionary fluctuation response relationship, (b) the proportionality between genetic variance ($V_g$) and isogenic phenotypic variation ($V_{ip}$) throughout evolution, and (c) the reduction of phenotypic dimensions to one or a few dimensions after evolution, in the following examples:

(I) Reaction-network dynamics model: As already mentioned, from the evolution simulation of a few toy cell models with a catalytic reaction network, the aforementioned points are confirmed, by adopting the fitness condition to increase the cell-growth speed or to enhance concentration of certain components\cite{CFKK-JTB,CFKK-PRE,Sato-KK1}.

(II) Gene regulation network model: Gene expressions, involving mRNA and protein synthesis, form a network governed by mutual activation or inhibition. The dynamics of gene expression levels follow threshold-type on-off dynamics, characterized by a matrix of positive and negative elements for activation and inhibition, akin to neural activities within neural networks. 
By assigning fitness to specific expression pattern for target genes and evolving the network, we have again verified properties (a)-(c), particularly when introducing a substantial level of noise\cite{KK-PLoS,KK-ESB}.
Furthermore, the reduced dimensionality may encompass multiple dimensions if various expression patterns correspond to different external conditions\cite{Sato-KK2023}.

(III) Evolving-interacting-spin model analogous to a spin-glass model: The above dynamics of gene regulation networks share similarities with spin dynamics ($S_i$) influenced by ferro (activating) and anti-ferro (inhibitory) interactions. By formulating a Hamiltonian ($H=\sum_j J_{ij}S_iS_j$) with an interaction matrix ($J_{ij}$), the distribution of spin configurations under temperature ($T$) is given by $exp(-H/(kT))$. From the configuration, the fitness is determined, for instance, by the overlap with a predetermined specific configuration of target spins. Accordingly, the matrix $J_{ij}$ undergoes evolution through mutation and selection. It is found that fitted states with robustness to mutation are evolved for a certain temperature range\cite{Sakata-Hukushima-KK}, whereas we observe properties (a)-(c) within a certain temperature range\cite{Sakata-KK,Pham-KK1}, indicative of robustness to mutation. This robustness is observed only at the replica symmetric phase, whereas for lower temperature, the robustness is collapsed due to replica symmetry breaking. 

(IV) Evolution of protein structure: Through a combination of data analysis, elastic-network models, and theoretical analysis, we have confirmed the tight correlation between $V_g$ and $V_{ip}$ as well as the dimensional reduction of protein configuration dynamics. Estimated configuration changes resulting from noise and genetic mutations are constrained within a common low-dimensional space, typically around five dimensions, a stark reduction from the hundreds of residues giving rise to the high-dimensional state space \cite{Tang-KK}. Note that such dimension reduction from high-dimensional state space is also discussed in protein dynamics\cite{Tlusty}.

Recent experimental research involving the evolution of bacteria exposed to a hundred antibiotics by Furusawa's group has provided further support. Transcriptome analysis of thousands of mRNA expression data indicates that changes in bacterial responses can be predicted by approximately seven principal components \cite{Furusawa}.
The dimensional reduction has also gathered significant attention in other fields of biology, as has been explored experimentally in development of {\sl C. elegans}\cite{Jordan}, laboratory ecological evolution \cite{Leibler}, and neural learning process in monkey\cite{Batista}, whereas insensitivity of phenotype to most parameters is discussed as sloppy parameter hypothesis\cite{sloppy}. 

\section{Towards statistical physics theory for evolutionary dimensional reduction}

While the argument for dimensional reduction grounded in evolutionary robustness and plasticity appears plausible, it remains somewhat speculative in the absence of a rigorous theoretical foundation. Theory to establish the macro-micro consistency is missing at present. Drawing inspiration from established theories in statistical mechanics, several avenues of investigation are expected:

(i) {\bf Projection to dominant collective mode from high-dimensional phenotype space}: In statistical physics, the method of projection onto the collective mode is established by Mori and Zwanzig \cite{Mori,Zwanzig} to be consistent with thermodynamics. In contrast, in the context of evolutionary dimensional reduction, the space onto which the projection is made emerges as a consequence of evolution itself. Consequently, we need to introduce a self-consistent approach to this projection method. One promising direction will be the marriage of dynamical mean-field theory \cite{Sompolinsky,Opper} with evolutionary dynamics \cite{Pham-KK2}, although the formulation of the dimension reduction process remains an ongoing endeavor.

(ii) {\bf Renormalization group}: Through evolutionary process to achieve robustness, microscopic perturbations are smeared out, leaving behind only the low-dimensional aspects of plasticity. This process bears resemblance to the elimination of irrelevant variables in the renormalization group \cite{Kadanoff, Wilson, Goldenfeld, Oono}. Devising a methodology to integrate renormalization group concepts with evolutionary processes remains a challenging frontier.

(iii) {\bf Slow-fast symmetry breaking}: As uncovered in the analysis of eigenvalue spectrum, few slow modes are separated from other many degrees of freedom, as represented by few outliers (whose eigenvalues are close to zero). As discussed, this separation of few slow modes is expected to facilitate the evolution. This phenomenon bears similarity to the origin of the central dogma, wherein symmetry breaking leads to the separation of information molecules with slower changes and functional molecules exhibiting faster changes \cite{Takeuchi-KK}. A similar separation phenomenon also emerges in neural systems, where separation of area with slower neural activities is relevant to Bayesian inference \cite{Ichikawa-KK}. Theoretical framework of slow-fast symmetry breaking for evolution is needed.

(iv) {\bf Avoidance of replica symmetry breaking}: In the evolving interacting spin model we mentioned in the last section, robustness to noise and dimensional reduction is achieved at the replica symmetric phase, whereas the replica symmetry breaking (RSB) will undermine robustness. In the case of evolving interaction, the applicability of replica theory extends to both spins and couplings\cite{Pham-KK1}. On the other hand, when the evolution under multiple fitness against different environmental conditions is postulated, the region of fitted replica symmetric phase shrinks towards RSB side\cite{Sakata-KK2023}, which may suggest that the state achieving both robustness and plasticity to multiple conditions will approach "the edge of glass". This intriguing phenomenon warrants further investigation. 

(v){\bf Macroscopic potential for cellular state}:
If one assumes dimensional reduction, it raises the possibility of a macroscopic potential theory akin to thermodynamics. Analogous to thermodynamics where free-energy potential emerges from dimensional reduction, one might envision a similar description of biological systems employing a limited set of macroscopic variables. This could entail the formulation of a macroscopic potential, in which the growth rate ($\mu_g$) or fitness is expressed as a function of environmental and genetic changes, as $\mu_g(E,G)$ \cite{KKCF-Rev}. This framework has the potential to yield linear relationships between evolutionary responses and phenotypic fluctuations caused by noise, as corroborated by bacterial evolution experiments, which were discussed in the previous section. In fact, landscape picture akin to potential was proposed by Waddington, with qualitative discussions on correlations between genetic and environmental responses, as is referred to {\sl genetic assimilation}\cite{Waddington}, a concept that resonates with the $V_g$-$V_{ip}$ relationship presented in this paper. With the formalization of the macroscopic potential, these discussions can advance to the level of quantitative theory.

\section{Constructing universal biology}

In the present paper, we discussed the premise of universal biology, by focusing on the consistency between slow evolutionary change and fast adaptive changes. Self-consistency principle between macro-micro scales is generally important. 

\subsection{Ideal cell model?}

At the cellular level, achieving consistency between the replication of numerous molecular species and cellular reproduction is essential. To comprehend this phenomenon, a simplified ideal cell model proves invaluable.
In the development of thermodynamics, the concept of an ideal gas played a pivotal role, even though thermodynamics itself is not limited to this particular case. Similarly, to facilitate the formulation of a macroscopic theory for cellular states, the introduction of an {\sl ideal cell model} capable of sustaining robust, steady growth may prove beneficial \cite{Zipf, SOC-Zipf, Crooks,Hashimoto, Jain}. In this context, potential state variables could include the growth rate, nutrient consumption, and the energetic costs associated with maintaining a cell, although additional macroscopic parameters might also be necessary. Establishing a possible connection with standard thermodynamic quantities, such as entropy production, becomes a crucial endeavor \cite{Himeoka, Dill}.

Furthermore, microbial cells typically undergo a transition from exponential growth to a stationary phase during which cellular growth ceases, and the cell enters a {\sl sleeping} state. Once a cell reaches this dormant state due to nutrient depletion, it generally requires a specific duration, referred to as the lag time, to resume growth upon the reintroduction of nutrients. Recent experimental observations have unveiled certain empirical relations among the lag time, starvation duration, and maximal growth rate \cite{Balaban}. By extending the ideal cell model to incorporate interactions with components that do not contribute to growth or catalysis, the transition to the sleeping state, as well as the aforementioned, empirical laws have recently been derived \cite{Himeoka-PRX}. This endeavor will be akin to the extension of the ideal gas model to the van der Waals model.

\subsection{Multicellularity and irreversible differentiation}

\begin{table}[ht]
\begin{center}
\begin{tabular}{|c|c|}
\hline
Thermodynamics & Universal Biology\\
\hline
Restriction to Eqb. Systems & Restriction to Steady Growth \\
\hline
Stability of Eqb. State & Robustness of Evolved State \\
\hline
Irreversibility to Eqb. & Return to Adapted State \\
\hline
Irreversiblity between Eqb. States & Differentiation from Stem Cells\\
\hline
Fluctuation-response Relation& Evolutionary Fluctuation-Response \\
\hline
Le Chatelier principle& Homeostasis by Adaptation\\
\hline
Thermodynamic Potential & Growth Macroscopic Potential \\
\hline
Free Energy vs Energy & Growth versus Nutrient Uptake \\
\hline
Ideal Gas & Ideal Cell Model for Growth \\
\hline
Liquid-Gas Transition & Exponential-Stationary Transition \\
\hline
\end{tabular}
\caption{Possible correspondence between thermodynamics and macroscopic phenomenological theory for biological system. Eqb is the abbreviation of equilibrium.}
\end{center}
\end{table}

We have hitherto explored the prospect of a macroscopic description for a cell. It is imperative to acknowledge that biological systems encompass a hierarchical structure, spanning from molecules to cells, and further to multicellular organisms, culminating in ecosystems. The resilience of biological systems hinges on their ability to uphold consistency across these diverse levels \cite{KK-book}. In multicellular organisms, the cellular community, be it a tissue or an entire organism, is intricately developed to ensure that both the collective ensemble of cells and each individual intra-cellular state remain robust in the face of perturbations.
In this case, cells differentiate into distinct types, with the relative abundance of each cell type maintained within a certain range.
 
The dynamical-systems approach for multiple cellular states finds its origins in the pioneering work of Goodwin \cite{Goodwin} and Kauffman \cite{Kauffman}. However, comprehending robustness at the multicellular level, as well as the irreversible loss of multipotency (the capacity to generate various cell types during differentiation from embryonic stem cells), necessitates an intricate interplay between intra-cellular dynamics and cell-cell interactions \cite{Furusawa-KK-Science}. It prompts the question: Can we characterize irreversibility and derive governing principles through the formulation of a macroscopic phenomenological theory? This stands as the next challenge once a phenomenological theory for a cell is established.

To sum up, we have discussed the possibility of macroscopic universal biology, inspired by thermodynamics, as summarized in the Table I.

This research was partially supported by and the Novo Nordisk Foundation (0065542) and Grant-in-Aid for Scientific Research (A) (20H00123) from the Ministry of Education, Culture, Sports, Science and Technology (MEXT) of Japan. The author would like to thank Chikara Furusawa, Tuan Pham, Takuya Sato, Ayaka Sakata, Qian-Yuan Tang, Tetsuhiro S. Hatakeyama, and Yuichi Wakamoto for useful discussions.

\end{document}